\documentstyle[prl,multicol,aps,epsfig,amstex]{revtex}

\begin{document}

\draft

\title{Raman scattering study of anomalous spin-, charge-, and lattice-dynamics
in the charge-ordered phase of ${\bf Bi}_{1-x}{\bf Ca}_x{\bf MnO}_3$
($x>0.5$)}

\author{S. Yoon$\rm ^{1}$, M. R\"ubhausen$\rm ^{1}$, S. L. Cooper$\rm ^{1}$,
K. H. Kim$\rm^{2,3}$, and S-W. Cheong$\rm ^{2,4}$}
\address{$\rm ^1$Department of Physics and Frederick Seitz Materials
Research Laboratory,\\ University of Illinois at Urbana-Champaign,
Urbana, IL 61801}
\address{$\rm ^2$Department of Physics and Astronomy, Rutgers University,
Piscataway, NJ 08854}
\address{$\rm ^3$Department of Physics, Seoul National University, Seoul
151-742, Korea}
\address{$\rm ^4$Bell Laboratories, Lucent Technologies, Murray Hill,
NJ 07974}

\date{\today}

\maketitle

\begin{abstract}
We report an inelastic light scattering study of the effects of
charge-ordering on the spin-, charge-, and lattice-dynamics in ${\rm
Bi}_{1-x}{\rm Ca}_{x}{\rm MnO}_3$ $(x>0.5)$. We find that
charge-ordering results in anomalous phonon behavior, such as the
appearance of `activated' modes. More significantly, however, the
transition to the CO phase results in the appearance of a
quasielastic scattering response with the symmetry of the spin
chirality operator ($T_{1g}$); this scattering response is thus
indicative of magnetic or chiral spin fluctuations in the AFM
charge-ordered phase.
\end{abstract}


\begin{multicols}{2}
\narrowtext

% main

Among the most interesting and rich phenomena exhibited by complex
transition metal oxides such as the nickelates\cite{nick},
cuprates\cite{tranquada}, and manganites\cite{chen} is charge- and
orbital-ordering, i.e., the organization of charges and orbital
configurations in periodic arrays on the lattice. The considerable
recent effort devoted to understanding this behavior has revealed a
variety of interesting properties, including novel states of matter
such as coexisting magnetic phases\cite{Liu,Dagotto} and possible
`quantum liquid crystal' states\cite{kivelson}. Yet, a number of
important issues remain unsolved, including the effects of orbital
and charge ordering (CO) on the lattice and charge dynamics, and the
nature of carrier motion in the complex spin background of the
N\'{e}el state. A clarification of these issues demands experimental
methods capable of probing the strong interplay among the spin-,
charge-, lattice-, and orbital-degrees-of-freedom in
strongly-correlated systems.

In this Letter, we discuss an inelastic light (Raman) scattering
study of the unconventional lattice-, spin-, and charge-dynamics in
the CO phase of the ${\rm Bi}_{1-x}{\rm Ca}_{x}{\rm MnO}_3$
$(x>0.5)$ system. Raman scattering offers several unique features in
the investigation of charge-ordered systems. For example, by
providing energy, symmetry, and lifetime information concerning
lattice-, spin-, as well as charge-excitations, Raman scattering
affords unique insight into the interplay among these coupled
excitations in various phases. Also, as a technique that can
sensitively probe unconventional charge- and spin-dynamics, such as
exotic ``chiral" spin and charge currents\cite{a2g,lee,shastry},
Raman scattering offers a unique means of probing the unconventional
spin- and charge-dynamics that arise when charge-carriers are placed
in the complex spin environment of CO systems.

These benefits are clearly evident in the present study, which
uncovers several interesting features of CO behavior in ${\rm
Bi}_{1-x}{\rm Ca}_{x}{\rm MnO}_3$ ($x>0.5$). First, polarized Raman
measurements show that charge-ordering results in the appearance of
activated phonon modes, due to the lowering of symmetry by
charge-stripe formation. Most interesting, however, is the
observation that a quasielastic Raman scattering response, with the
symmetry of the spin chirality operator ($T_{1g}$), develops in the
CO phase. This distinctive scattering response indicates the
presence of chiral fluctuations at finite temperatures in the CO/AFM
phase, possibly arising from a chiral spin-liquid state associated
with the Mn core spins, or from closed-loop charge motion caused by
the constraining environment of the complex orbital and N\'{e}el
spin textures.

The samples used in our study were flux-grown single crystalline
$\rm Bi_{0.19}Ca_{0.81}MnO_3$ ($ T_{\rm co} = 165$ K, $T_{\rm N} =
120$ K) and $\rm Bi_{0.18}Ca_{0.82}MnO_3$ ($ T_{\rm co} = 210$ K,
$T_{\rm N} = 160$ K). The typical dimensions of these samples are 2
x 2 x 1 mm$^3$. Raman spectra were measured in a backscattering
geometry using continuous helium flow and cold-finger optical
cryostats, and a modified subtractive-triple-grating spectrometer
equipped with a nitrogen-cooled CCD array detector. The spectra were
corrected for the spectral response of the spectrometer and the
detector. The samples were excited with 4 mW of the 4762-$\rm \AA$
line of the Kr$^+$ laser, focused to a 50 $\mu$m diameter spot
within a single CO domain of the crystals. Temperatures listed for
the Raman spectra include estimates of laser heating effects. To
identify excitation symmetries, the spectra were obtained with the
incident ($\mathbf{E_i}$) and scattered ($\mathbf{E_s}$) light
polarized in various configurations, including ($\rm \bf E_{i},
E_{s}$) = ($\mathbf{x}$,$\mathbf{x}$) and
($\mathbf{y}$,$\mathbf{y}$): $A_{1g}+E_g$, and ($\rm \bf E_{i},
E_{s}$) = ($\mathbf{L}$,$\mathbf{L}$):
$A_{1g}+\frac{1}{4}E_g+T_{1g}$, where x and y are the [100] and
[010] crystal directions, respectively, L is left circular
polarization, and where $A_{1g}$, $E_g$ and $T_{1g}$ are
respectively the singly-, doubly-, and triply-degenerate irreducible
representations of the $O_h$ space group of the crystals, which have
a pseudocubic structure\cite{MT,Mori,rada,Bao}.

\begin{figure}[htbp]
    \centering
    \leavevmode
        \epsfig{file=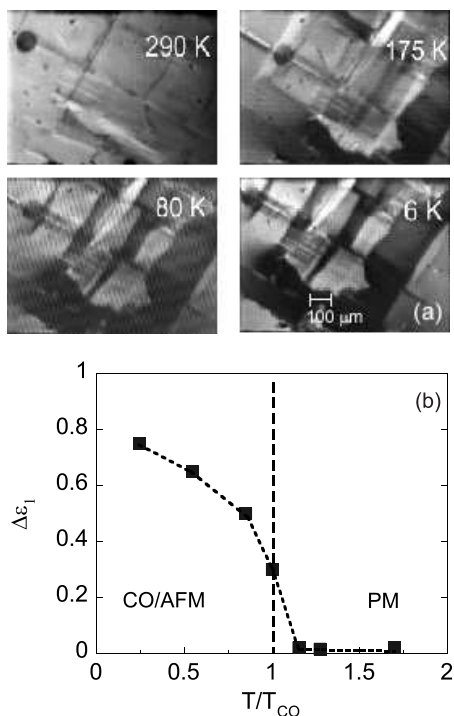,width=6cm}
    \caption{(a) Polarized microscope images of developing CO domains
    at various temperatures. (b) Dielectric anisotropy parameter
    $\Delta^{\epsilon_1}$ (see text) at $\omega = 1$ eV.
    The dotted line is a guide to the eye.}
\label{Fig1}
\end{figure}

Figure \ref{Fig1} (a) shows polarized microscope images of the
\noindent (100) $\rm Bi_{0.19}Ca_{0.81}MnO_3$ sample surface taken
at 290 K, 175 K, 80 K, and 6 K, respectively. One can clearly see
the growth of ``light" and ``dark" regions below the charge-ordering
temperature, $T_{\rm co} = 165$ K, corresponding to the development
of domains having perpendicular orientations of the charge-stripes.
The evolution of CO behavior and domain formation below $T_{\rm co}$
is more quantitatively illustrated in Fig.~\ref{Fig1} (b), which
presents the temperature dependence of the dielectric anisotropy,
$\Delta^{\epsilon_1}(\omega) =
\frac{|\epsilon_1^{ac}-\epsilon_1^{bc}|}{\sqrt{
(\epsilon_1^{ac})^2+(\epsilon_1^{bc})^2}}$, where $\epsilon_1^{ac}$
and $\epsilon_1^{bc}$ are the dielectric responses for $\omega = 1$
eV light polarized in the ac and bc planes respectively\cite{rub1}.
Notably, the temperature-dependence of $\Delta^{\epsilon_1}$ is
similar to that of the order-parameter in a second-order phase
transition, consistent with our expectation that the increasing size
of this quantity below $T_{\rm co}$ reflects the increasing
organization of charges below $T_{\rm co}$. However, at low
temperatures, $\Delta^{\epsilon_1}$ saturates below the maximum
value of 1 due to the fact that the optical spot in the ellipsometry
measurements is not isolated to a single domain.

\begin{figure}[htbp]
    \centering
    \leavevmode
        \epsfig{file=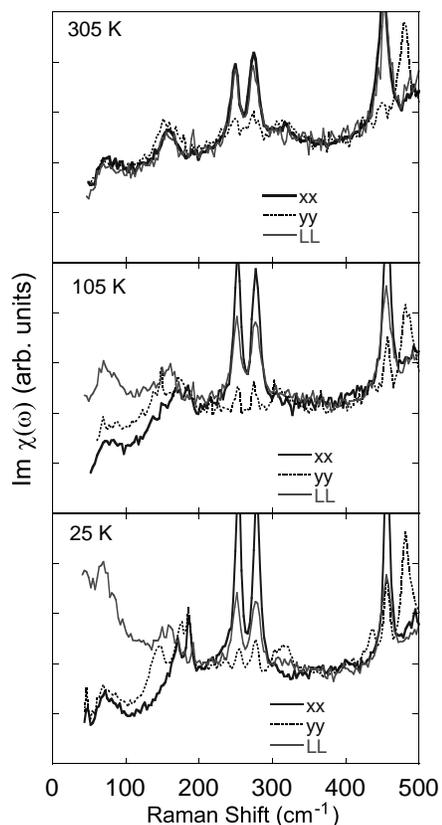,width=5.7cm}
    \caption{Raman spectra of $\rm Bi_{0.19}Ca_{0.81}MnO_3$
    in xx, yy and LL polarization configurations at 305 K,
    105 K, and 25 K.}
\label{Fig2}
\end{figure}

One of the distinct advantages of polarized Raman scattering
techniques for measuring optical anisotropy in the CO phase is that
this technique allows the study of single ($\lesssim 100$ $\mu$m)
domains with uniformly aligned charge stripes. Raman spectra of
single domain regions in $\rm Bi_{0.19}Ca_{0.81}MnO_3$ are
illustrated, for various temperatures and scattering geometries, in
Fig.~\ref{Fig2}. Note first that in the high temperature
``isotropic" phase ($T = 305$ K),
\noindent all three scattering geometries (xx, yy, and LL) overlap,
with the exception of some intensity differences associated with the
phonons. However, with decreasing temperature into the CO phase, two
significant features are evident: First, several changes in the
phonon spectra evolve with decreasing temperature, including the
appearance of new modes and the evolution of differences in the
phonon spectra associated with the xx and yy scattering geometries.
Second, while the low frequency backgrounds associated with the xx
and yy scattering geometries decrease with decreasing temperature
into the CO phase, there is a dramatic growth of the low frequency
scattering background in the LL geometry, betraying the development
of a distinctive $T_{1g}$-symmetry quasielastic scattering response
in the CO/AFM phase.

\begin{figure}[htbp]
    \centering
    \leavevmode
        \epsfig{file=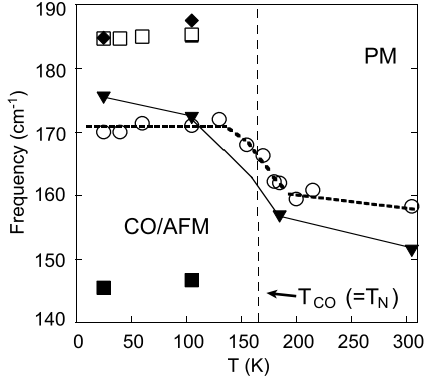,width=5.6cm}
    \caption{Hardening of the Mn vibration mode ($\sim 160$
    cm$^{-1}$) and the development of additional modes ($\sim 145$
    cm$^{-1}$ and $\sim 185$ cm$^{-1}$) below the CO transition.
    Open symbols denote data from xx, and filled symbols denote
    data from yy scattering geometries. The lines are guides to
    the eye.}
\label{Fig3}
\end{figure}

We focus first on the effects of charge-ordering on the phonons in
$\rm Bi_{0.19}Ca_{0.81}MnO_3$ - such information is important, as
the optical phonons function as `local probes' of changes in the
local symmetry and bond-strengths caused by
charge-ordering\cite{yama}. Consider first the $\sim 160$ cm$^{-1}$
$A_{1g}$ phonon mode in Fig.~\ref{Fig3} (circles and triangles),
which is associated with in-phase Mn vibrations. This mode exhibits
an abrupt hardening across the charge-ordering transition,
indicative of the effects of charge-ordering on the lattice force
constants via changes in the Coulomb energies. Also,
Figs.~\ref{Fig2} and ~\ref{Fig3} illustrate the appearance of a
second mode at $\sim 185$ cm$^{-1}$. In the
\noindent isotropic high temperature phase ($T>T_{\rm co}$), this
mode is present in the xy scattering configuration, but develops
also in the xx scattering geometry below $T_{\rm co}$ due to the
breakdown of symmetry selection rules in the CO phase.

The breaking of 4-fold in-plane symmetry due to long-range
charge-ordering is also reflected in differences in the phonon
spectra observed in the xx and yy scattering geometries. In
particular, the yy spectrum at 25 K (Fig.~\ref{Fig2}) shows the
development of `new' phonon modes near $\sim 145$ cm$^{-1}$ (filled
squares in Fig.~\ref{Fig3}), $\sim 300$ cm$^{-1}$, and $\sim 420$
cm$^{-1}$. The appearance of new modes can reflect
Brillouin-zone-folding of zone boundary modes to the zone center,
caused by the additional periodicity associated with charge-stripe
formation. It is expected, however, that zone-folded modes should
have much weaker intensities than `regular'
modes\cite{sherman,lemm}, which is not the present case. A more
plausible interpretation therefore is that the new modes we observe
are ``activated" modes due to charge-ordering: For example, the
out-of-phase Mn vibrational mode is not Raman-active in the PM
`isotropic' phase because the net charge fluctuation associated with
out-of-phase motion of the (${\rm Mn^{3.5+}-O-Mn^{3.5+}}$) complex
is zero, and hence this mode cannot modulate the polarizability.
However, in the CO phase, the different charges on the Mn$^{3+}$ and
Mn$^{4+}$ sites cause out-of-phase Mn vibrations of the (${\rm
Mn^{3+}-O-Mn^{4+}}$) complex to have a non-zero net charge
fluctuation that couples to the polarizability, resulting in the
`new' Raman-active mode at $145$ cm$^{-1}$.

\begin{figure}[htbp]
    \centering
    \leavevmode
        \epsfig{file=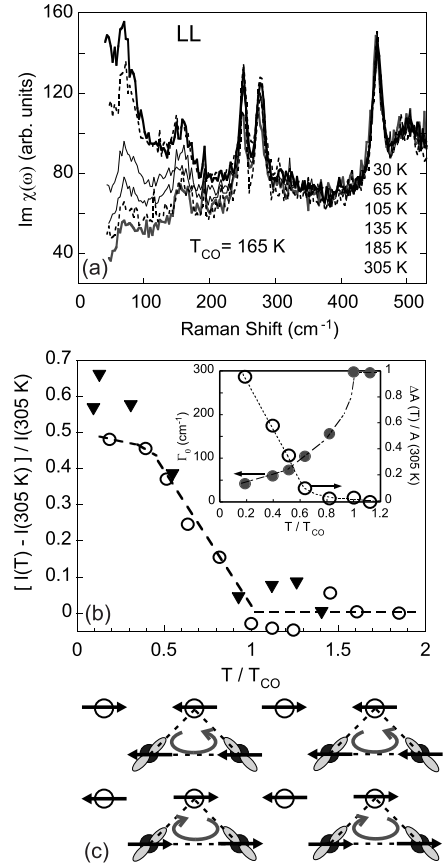,width=5.8cm}
    \caption{(a) Temperature dependence of the Raman spectra in the
    LL scattering geometry. (b) Fractional change in the integrated
    quasielastic Raman scattering intensity ($50 - 350$ cm$^{-1}$) as
    a function of temperature for the $T_{\rm co} = 165$ K (circles)
    and $T_{\rm co} = 210$ K (triangles) samples.
    Inset: Fractional change in the quasielastic scattering
    amplitude A and fluctuation rate $\Gamma$ as a function of
    temperature. Lines are guides to the eye. (c) Example of a
    closed-loop path for charge motion in the CO phase ($x=0.5$)
    which is not precluded by either the orbital configuration or
    by the spin environment. Filled and empty circles represent
    ${\rm Mn^{3+}}$ and ${\rm Mn^{4+}}$ sites, respectively.}
\label{Fig4}
\end{figure}

An even more interesting feature apparent in Figs.~\ref{Fig2} and
~\ref{Fig4} (a) is the development in the CO state of a quasielastic
Raman response in the LL scattering geometry,
\begin{equation}
    {\rm Im} \chi(\omega) \sim
    \frac{A \omega \Gamma}{\omega^2 + \Gamma^2}
\end{equation}
where A is the quasielastic scattering amplitude and $\Gamma$ is the
fluctuation rate. The absence of a similar scattering
\noindent response in either xx or yy scattering geometries
definitively identifies this quasielastic response as having
$T_{1g}$ symmetry. This distinctive scattering symmetry transforms
like the spin-chirality operator ($\vec{S_1} \cdot \vec{S_2} \times
\vec{S_3}$)\cite{shastry}, and is typical of scattering from
magnetic fluctuations\cite{cooper} and from chiral spin
fluctuations\cite{shastry}. Thus, while the ground state of the
CO/AFM phase is not generally expected to have a net magnetization
or spin chirality, the development of this $T_{1g}$ quasielastic
response in Figs.~\ref{Fig2} and ~\ref{Fig4} (a) betrays the
presence of strong {\it fluctuations} associated with such a broken
time-reversal symmetry state at finite temperatures in the CO/AFM
phase. Interestingly, the fluctuation rate $\Gamma$ associated with
this unusual response (inset Fig.~\ref{Fig4} (b)) tends to zero with
decreasing temperature, perhaps indicating a tendency toward static
long-range order as $T \rightarrow 0$.

When considering the origin of this anomalous scattering response,
we can first rule out a ``precursor" fluctuational response
associated with either charge-stripe- or orbital-ordering: while
such responses should develop above, become maximum near, and
diminish below the ordering transition temperature $T_{\rm co}$,
Figs.~\ref{Fig4} (a) and (b) clearly illustrate that the $T_{1g}$
quasielastic response we observe evolves at $T_{\rm co}$, and grows
with decreasing temperature below $T_{\rm co}$, coincident with the
charge-order-parameter $\Delta^{\epsilon_1}$ in Fig.~\ref{Fig1} (b).

Several intriguing possibilities are consistent with both the
distinctive $T_{1g}$ symmetry and temperature dependence of the
quasielastic light scattering response in Figs.~\ref{Fig4} (a) and
(b). First, although neutron scattering studies show no evidence for
ferromagnetic spin fluctuations below $T_{\rm N}$ in this
system\cite{Bao}, it is possible that the $T_{1g}$ scattering we
observe reflects spin fluctuations associated with a canted AFM
phase. Similarly, the properties of the quasielastic scattering
response in Fig.~\ref{Fig4} are also consistent with the presence of
chiral spin fluctuations associated with the core spins, for example
similar to those observed in ferromagnetic pyrochlores such as $\rm
Sm_{2}Mo_{2}O_{7}$\cite{taguchi}. Such fluctuations of canted AFM or
spin chirality could arise in the CO manganites due to geometrical
frustration of the Mn core spins, and/or to an appreciable
Dzyaloshinskii-Moriya interaction ($\sim \vec{S_1} \times
\vec{S_2}$), in the AFM/CO phase.

Finally, another interesting possibility is that the fluctuational
response in Fig.~\ref{Fig4} is associated with chiral charge
currents, i.e., charges constrained to hop in closed-loop paths. The
possibility of such charge motion in the N\'{e}el spin environment
of the CO/AFM phase is suggested by first noting that long-range
translational charge motion is constrained in this phase by the
complex orbital and N\'{e}el spin structure, by the constraints of
the double-exchange hopping mechanism, and by disorder (e.g., by
doping away from commensurate fillings), which strongly limits
conduction along the 1D Mn$^{3+}$ chains\cite{maez}. However,
Fig.~\ref{Fig4} (c) illustrates one possible closed-loop path in
which the hopping of holes is not constrained by either the spin or
orbital environments in the CO/AFM phase. Interestingly, Nagaosa and
Lee predicted that such ``closed-loop" charge hopping should be
present in doped AFM insulators, and should be manifest in the
appearance of a quasielastic Raman response\cite{lee} similar to
that observed in the CO/AFM phase of ${\rm Bi}_{1-x}{\rm Ca}_{x}{\rm
MnO}_3$ ($x>0.5$) (Fig.~\ref{Fig4} (a)). Such quasielastic light
scattering arises in this case from fluctuations in an induced
effective magnetic field generated by the chiral charge currents,
$\langle \chi(m) \chi^{\dagger}(m) \rangle \sim \langle m
m^{\dagger} \rangle$, where $\chi(m)$ is the `field'-dependent
electric susceptibility.

In conclusion, our Raman scattering studies of ${\rm Bi}_{1-x}{\rm
Ca}_{x}{\rm MnO}_3$ have allowed us to explore the influence of
charge- and orbital-ordering on the lattice-, charge-, and
spin-dynamics. Most significantly, in the CO/AFM phase these studies
reveal the development of a fluctuational (quasielastic) response
with the distinctive symmetry of the spin-chirality operator -- this
remarkable response is consistent with the presence of a fluctuating
chiral state at finite temperatures in the CO/AFM state.
Importantly, these studies also clearly demonstrate that Raman
scattering is uniquely suited to probing exotic ``chiral" phases in
other correlated systems, such as the CMR-phase
manganites~\cite{chun}, geometrically-frustrated metallic
ferromagnets~\cite{taguchi}, the high $T_{c}$ cuprates~\cite{lee},
and quantum Hall systems.

We thank M. V. Klein, Y. Lyanda-Geller, and P. Goldbart for useful
discussions. We acknowledge financial support of the DOE via
DEFG02-96ER45439 (S.Y., M.R., S.L.C.), the DFG via Ru 773/1-1, the
NSF through the STCS via DMR91-20000 (M.R.), and NSF-DMR-9802513
(K.H.K., S-W.C.).

%\end{multicols}

%\widetext

\end{multicols}


\begin{references}

\bibitem{nick} C. H. Chen, S-W. Cheong, and A. S. Cooper,
Phys. Rev. Lett. {\bf 71}, 2461 (1993).

\bibitem{tranquada} J. M. Tranquada {\it et al.}, Nature (London) {\bf 375},
561 (1995).

\bibitem{chen} C. H. Chen and S-W. Cheong, Phys. Rev. Lett. {\bf 76}, 4042
(1996).

\bibitem{Liu} H. L. Liu, S. L. Cooper, and S-W. Cheong, Phys. Rev. Lett.
{\bf 81}, 4684 (1998).

\bibitem{Dagotto} A. Moreo, S. Yunoki, and E. Dagotto, Science {\bf 283},
2034 (1999).

\bibitem{kivelson} S. A. Kivelson, E. Fradkin, and V. J. Emery,
Nature (London) {\bf 393}, 550 (1998).

\bibitem{a2g} D. V. Khveshchenko and P. B. Wiegmann, Phys. Rev.
Lett. {\bf 73}, 500 (1994).

\bibitem{lee} N. Nagaosa and P. A. Lee, Phys. Rev. B {\bf 43},
1233 (1991).

\bibitem{shastry} B. S. Shastry and B. I. Shraiman, Phys. Rev.
Lett. {\bf 65}, 1068 (1990); P. E. Sulewski {\it et al.}, {\it
ibid.}, {\bf 67}, 3864 (1991).

\bibitem{MT} Y. Moritomo {\it et al.}, Phys. Rev. B {\bf 55}, 7549
(1997).

\bibitem{Mori} S. Mori, C. H. Chen, and S-W. Cheong, Nature
(London) {\bf 392}, 473 (1998).

\bibitem{rada} P. G. Radaelli {\it et al.}, Phys. Rev. B {\bf 55},
3015 (1997).

\bibitem{Bao} W. Bao {\it et al.}, Phys. Rev. Lett. {\bf 78}, 543
(1997).

\bibitem{rub1} M. R\"ubhausen {\it et al.}, Phys. Rev. B {\bf 62},
R4782 (2000).

\bibitem{yama} K. Yamamoto {\it et al.}, J. Phys. Soc. Jpn {\bf
68}, 2538 (1999).

\bibitem{sherman} E. Ya. Sherman {\it et al.}, Europhys. Lett.
{\bf 48}, 648 (1999).

\bibitem{lemm} M. Fischer, {\it et al.}, Phys. Rev. B {\bf 60},
7284 (1999).

\bibitem{cooper} S. L. Cooper {\it et al.}, Phys. Rev. B {\bf 35},
2615 (1987); S. L. Cooper {\it et al.}, {\it ibid.}, {\bf 36},
5743 (1987).

\bibitem{taguchi} Y. Taguchi and Y. Tokura, Phys. Rev. B {\bf 60},
10280 (1999); K. Ohgushi {\it et al.}, cond-mat/9912206.

\bibitem{maez} R. Maezono, S. Ishihara, and N. Nagaosa,
Phys. Rev. B {\bf 57}, R13993 (1998).

\bibitem{chun} S. H. Chun {\it et al.}, Phys. Rev. Lett. {\bf 84},
757 (2000); Y. Lyanda-Geller {\it et al.}, cond-matt/9904331; J.
Ye {\it et al.}, Phys. Rev. Lett. {\bf 83}, 3737 (1999).

\end{references}
\end{document}